\documentclass{article}
\usepackage{graphicx}
\usepackage{amsmath}
\usepackage[toc,page]{appendix}
\usepackage{bbm} 
\usepackage{pdfpages} 
\usepackage{amssymb} 
\usepackage{multirow} 
\usepackage{bm} 
\usepackage{booktabs}
\usepackage{color}
\usepackage{array}
\begin{document}

\title{Explaining predictive models using Shapley values and non-parametric vine copulas}
%
%
\author{Kjersti Aas \and Thomas Nagler \and Martin Jullum \and Anders L\o land}
%
%

%
\maketitle 
\begin{abstract}
  The original development of Shapley values for prediction explanation relied on the assumption that the features being described were independent. If the features in reality are dependent
  this may lead to incorrect explanations. Hence, there have recently been attempts of appropriately modelling/estimating the dependence between the features. Although the proposed
  methods clearly outperform the traditional approach assuming independence, they have their weaknesses. In this paper we propose two new approaches for modelling the dependence between the features.
  Both approaches are based on vine copulas, which are flexible tools for modelling multivariate non-Gaussian distributions able to characterise a wide range of complex dependencies.
  The performance of the proposed methods is evaluated on simulated data sets and a real data set. The experiments demonstrate that the vine copula approaches give more accurate approximations to
  the true Shapley values than its competitors.
\end{abstract}
\section{Introduction}\label{sec:intro}

In many applications complex machine learning models are outperforming traditional regression models. It is often hard to understand why the machine learning models perform so well. Hence, during the last few years, a new line
of research has emerged focusing on interpreting the predictions from these models. One such method that has become very popular is Shapley values \cite{Lundberg,Strumbelj,Strumbelj2, aas2019Explaining}.
This method, which is based on concepts from cooperative game theory, was originally invented for assigning payout to players depending on their contribution towards the total payout \cite{Shapley53}.
When interpreting machine learning models, the model features are the players and the prediction is the total payout. 

The original development of Shapley values for prediction explanation \cite{Lundberg,Strumbelj,Strumbelj2} relied on the assumption that the features being described were independent. \cite{aas2019Explaining}
showed that if there is a high degree of dependence among some or all the features, this may lead to severely inaccurate Shapley value estimates and incorrect explanations. In the same paper, the authors deal with this problem, proposing
three different approaches appropriately modelling/estimating the dependence between the features; the Gaussian approach, the Gaussian copula approach,
and the empirical approach. Although all three methods clearly outperform the traditional approach assuming independence, they have their weaknesses. The Gaussian approach
assumes that features are multivariate Gaussian distributed, while the Gaussian copula approach represents the marginal distributions of the features with their empirical margins and model
the dependence structure by a Gaussian copula \cite{Joe96}. Hence, they will work well if respectively the distribution or the dependependy structure of the features is Gaussian.
The empirical approach is inspired by the kernel estimator. Like most other non-parametric density estimation approaches, this method suffers from the curse of dimensionality. It would therefore require a large data set to be
accurate in problems with many features.

In this paper, we propose two alternative approaches to estimate Shapley values. In both approaches, the multivariate joint density function of the features is represented by a vine copula \cite{Joe96}, but they differ
in the way the Shapley contribution function is evaluated. A vine copula is a multivariate copula that is constructed from a set of bivariate ones, so-called \textit{pair-copulas}. All of these bivariate copulas may be selected completely freely,
meaning that vine copulas are able to characterise a wide range of complex dependencies. Hence, the new approaches are expected to outperform the existing ones in cases where the feature distribution is far from the Gaussian. 

The main part of the methodology proposed in this paper may be used for many other applications than computing Shapley values. It may e.g. be regarded as a contribution to the field of non-parametric conditional density estimation.

The rest of the paper is organized as follows. We begin by explaining the fundamentals of the Shapley value framework in an explanation setting in Section \ref{sec:Shapley}, while Section \ref{prevMet} reviews some of of the previously
proposed Shapley methods for prediction explanation. In Section \ref{sec:nonParvines} we introduce the two new methods for computing Shapley values based on vine copulas. Section \ref{sec:simulation} presents various simulation studies
that demonstrate that our method works in a variety of settings, while Section \ref{sec:data} gives a real data example. Finally, in Section \ref{sec:conclusion}, we conclude.


\section{Shapley values}\label{sec:Shapley}

\subsection{Shapley values in game theory}\label{subsec:Shapley_framework}
Suppose we are in a cooperative game setting with $M$ players, $j = 1, \dots, M$, trying to maximize a payoff. Let $\mathcal{M}$ be the set of all players and $\mathcal{S}$ any subset of $\mathcal{M}$. Then the Shapley value \cite{Shapley53} for the $j$th player is defined as 


\begin{equation}\label{eq:shapley}
\phi_j = \sum_{\mathcal{S} \subseteq \mathcal{M} \backslash \{j\}} \frac{|\mathcal{S}|! (M - |\mathcal{S}| - 1)!}{M!} (v(\mathcal{S} \cup \{ j\}) - v(\mathcal{S})).
\end{equation}
Here, $v(\mathcal{S})$ is the contribution function which maps subsets of players to real numbers representing the worth or contribution of the group $\mathcal{S}$ and $|\mathcal{S}|$ is the number of players in subset $\mathcal{S}$. 

In the game theory sense, each player receives $\phi_j$ as their payout.  
From the formula, we see that this payout is just a weighted sum of the player's marginal contributions to each group $\mathcal{S}$. 
Lloyd Shapley  proved that distributing the total gains of the game in this way is `fair' in the sense that it obeys certain important axioms \cite{Shapley53}. 

\subsection{Shapley values for prediction explanation}\label{subsec:Shapley_explain}
In a machine learning setting, imagine a scenario where we $M$ features, $\bm{x} = (x_1, \dots, x_M)$ and a univariate response $y$, and have fitted the  model $g(\bm{x})$. We now want to explain the prediction $g(\bm{x}^*)$ for a
specific feature vector $\bm{x}^*$. The papers \cite{Lundberg,Strumbelj,Strumbelj2} suggest doing this with Shapley values where the predictive model replaces the cooperative game and the features replace the players.
To use \eqref{eq:shapley}, \cite{Lundberg} defines the contribution function $v(\mathcal{S})$ as the following expected prediction
\begin{equation}\label{eq:cond_expec}
v(\mathcal{S}) = \mathbb{E}[g(\bm{x}) | \bm{x}_\mathcal{S} = \bm{x}^*_\mathcal{S}]. 
\end{equation} 
Here $\bm{x}_\mathcal{S}$ denotes the features in subset $\mathcal{S}$ and $\bm{x}_\mathcal{S}^*$ is the subset $\mathcal{S}$ of the feature vector $\bm{x}^*$ that we want to explain. Thus, $v(\mathcal{S})$ denotes the expected
prediction given that the features in subset $\mathcal{S}$ take the value $\bm{x}_\mathcal{S}^*$.

If the features are continuous, we can write the conditional expectation \eqref{eq:cond_expec} as
\begin{equation}\label{eq:cond_integral}
\begin{split}
\mathbb{E}[g(\bm{x}) | \bm{x}_\mathcal{S} = \bm{x}_\mathcal{S}^*] &= \mathbb{E}[g(\bm{x}_{\bar{\mathcal{S}}}, \bm{x}_{\mathcal{S}})| \bm{x}_\mathcal{S} = \bm{x}_\mathcal{S}^*] = \int g(\bm{x}_{\bar{\mathcal{S}}}, \bm{x}_{\mathcal{S}}^*) f(\bm{x}_{\bar{\mathcal{S}}} | \bm{x}_{\mathcal{S}} = \bm{x}_{\mathcal{S}}^*) \,d\bm{x}_{\bar{\mathcal{S}}},
\end{split}
\end{equation} where $\bm{x}_{\bar{\mathcal{S}}}$ is the vector of features not in $\mathcal{S}$ and $f(\bm{x}_{\bar{\mathcal{S}}} | \bm{x}_{\mathcal{S}} = \bm{x}_{\mathcal{S}}^*)$ is the conditional density of $\bm{x}_{\bar{\mathcal{S}}}$ given $\bm{x}_{\mathcal{S}} = \bm{x}_{\mathcal{S}}^*$.
To compute Shapley values in practice, the conditional expecation in \eqref{eq:cond_integral} needs to be approximated empirically.
Note that in the rest of the paper we use lower case $x$-s for both random variables and realizations to keep the notation simple.

\section{Estimating Shapley values}\label{prevMet}

\subsection{The independence approach}
 
Since the conditional probability density is rarely known and difficult to estimate, \cite{Lundberg} replaces it with the simple (unconditional) probability density
\begin{equation}\label{eq:cond_is_marg}
f(\bm{x}_{\bar{\mathcal{S}}} | \bm{x}_{\mathcal{S}} = \bm{x}_{\mathcal{S}}^*) = f(\bm{x}_{\bar{\mathcal{S}}}).
\end{equation}
The integral is thus approximated by
\begin{equation}\label{eq:expectation_without_cond}
\mathbb{E}[g(\bm{x}) | \bm{x}_\mathcal{S} = \bm{x}_\mathcal{S}^*] \approx \int g(\bm{x}_{\bar{\mathcal{S}}}, \bm{x}_{\mathcal{S}}^*) f(\bm{x}_{\bar{\mathcal{S}}}) \,d\bm{x}_{\bar{\mathcal{S}}},
\end{equation}
which is estimated by randomly drawing $K$ times from the full training data set and calculating
\begin{equation}\label{eq:indep}
v_{\text{KerSHAP}}(\mathcal{S}) = \frac{1}{K} \sum_{k=1}^K g(\bm{x}_{\bar{\mathcal{S}}}^k, \bm{x}_\mathcal{S}^*).
\end{equation}
Here, $\bm{x}_{\bar{\mathcal{S}}}^k$, $k = 1, \dots, K$ are the samples from the training set and $g(\mathbf{\cdot})$ is the estimated prediction model. In the Shapley literature,
the approximation \eqref{eq:expectation_without_cond} is sometimes termed the interventional conditional expectation, while \eqref{eq:cond_integral} is denoted the observational conditional expectation.
See e.g. \cite{chen2020true} for more details.

Unfortunately, when the features are not independent, \cite{aas2019Explaining} demonstrates that naively replacing the conditional probability function with the unconditional one leads to very inaccurate
Shapley values.

\subsection{The Gaussian copula method}\label{GaussCop}

In \cite{aas2019Explaining}, one of the proposed methods for estimating $f(\bm{x}_{\bar{\mathcal{S}}} | \bm{x}_{\mathcal{S}} = \bm{x}_{\mathcal{S}}^*)$ without relying on the naive assumption of independence is
based on the \emph{Gaussian copula}. A copula is a function that characterizes the dependence in a random vector. By Sklar's theorem, any joint distribution function $F$ with marginal cdf's $F_1, \dots, F_M$ can be written as 
\begin{align*}
  F(\bm x) = C(F_1(x_1), \dots, F_M(x_M)),
\end{align*}
where $C$ is the copula function. Copulas are distribution functions with uniform margins. The corresponding density is denoted by $c$.

There are several parametric families for the copula function. The Gaussian copula is a special case. It is derived by inverting the above display i.e.,
\begin{align*}
  C(\bm u) = F(F_1^{-1}(u_1), \dots, F_M^{-1}(u_M)),
\end{align*}
and taking $F$ as a multivariate Gaussian distribution. This gives rise to a parametric model (parametrized by a correlation matrix) that reflects Gaussian dependence, but can be combined with arbitrary marginal distributions.

To compute Shapley values, we first need an estimate of the marginal distributions and copula parameters. \cite{aas2019Explaining} proposed to approximate the marginals $F_1, \dots, F_M$ by the corresponding empirical \emph{cdf}s and
parametrize the copula by the empirical correlation matrix of corresponding normal scores. Together this gives us an estimated model for the joint distribution
\[\hat F(\bm x) = \hat C(\hat F_1(x_1), \dots, \hat F_M(x_M)).\]
The conditional expectation in \eqref{eq:cond_integral} can now be approximated by simulating conditionally  from the estimated model. More precisely, let $\bm x_{{\bar{\mathcal{S}}}}^{k}, k = 1, \dots, K$ be simulated values
of $\bm x_{{\bar{\mathcal{S}}}}$ given $\bm x_{\mathcal{S}} = \bm x_{\mathcal{S}}^*$ and compute 
\begin{equation}\label{eq:montecarl}
  v_{\text{KerSHAP}}(\mathcal{S}) = \frac{1}{K} \sum_{k=1}^K g(\bm{x}_{\bar{\mathcal{S}}}^k, \bm{x}_\mathcal{S}^*). 
\end{equation}
Conditional simulation from the Gaussian copula can be achieved in essentially the same way as for the multivariate Gaussian distribution, see \cite{aas2019Explaining} for more details. 

The Gaussian copula model is very flexible with regard to the marginal distributions, but quite restrictive in the dependence structures it can capture. It can only represent radially symmetric dependence relationships and does not allow for tail
dependence (i.e., joint occurrence of extreme events has small probability). We therefore wish to use more flexible copula models and we shall focus on vine copula models specifically in what follows.

\section{Extending the Shapley framework with vine copulas}\label{sec:nonParvines}

A vine copula is a multivariate copula that is constructed from a set of bivariate ones, so-called \textit{pair-copulas}. All of these bivariate copulas may be selected completely freely as the 
resulting structure is guaranteed to be a valid copula. Hence, vine copulas are highly flexible being able to characterise a wide range of complex dependencies.

Vine copulas have become very popular over the last decade. The main idea was originally proposed by \cite{Joe96} and further explored and discussed by \cite{Bedford2,Bedford} and \cite{vinebook}. However, it was the
paper \cite{Aas07}, putting them in an inferential context, that really spurred a surge in empirical applications of these constructions. In this paper we use vine copulas to model the multivariate distributions
involved in the Shapley framework. After a brief introduction to vine copulas in Section \ref{subsec:vines}, we introduce two new methods for approximating the Shapely value contributions based on these structures in
Sections \ref{subsec:condsim} and \ref{subsec:ratio}. Finally, computationally efficient selection of the D-vine order is discussed in Section \ref{subsec:search}.

\subsection{Background on vine copulas}\label{subsec:vines}

In a vine copula is a multivariate copula the copula density is decomposed into a product of pair-copula densities. This decomposition is not unique. To organize all possible decompositions, the notion of {\it regular vines} (R-vines)
was introduced by \cite{Bedford}, 
and described in more detail in \cite{vinebook}. It involves the specification of a sequence of 
trees, each edge of which corresponds to a pair-copula. These pair-copulas constitute the building 
blocks of the joint R-vine distribution.

\begin{figure}[t]
 \centering
 \includegraphics[width=0.7\linewidth]{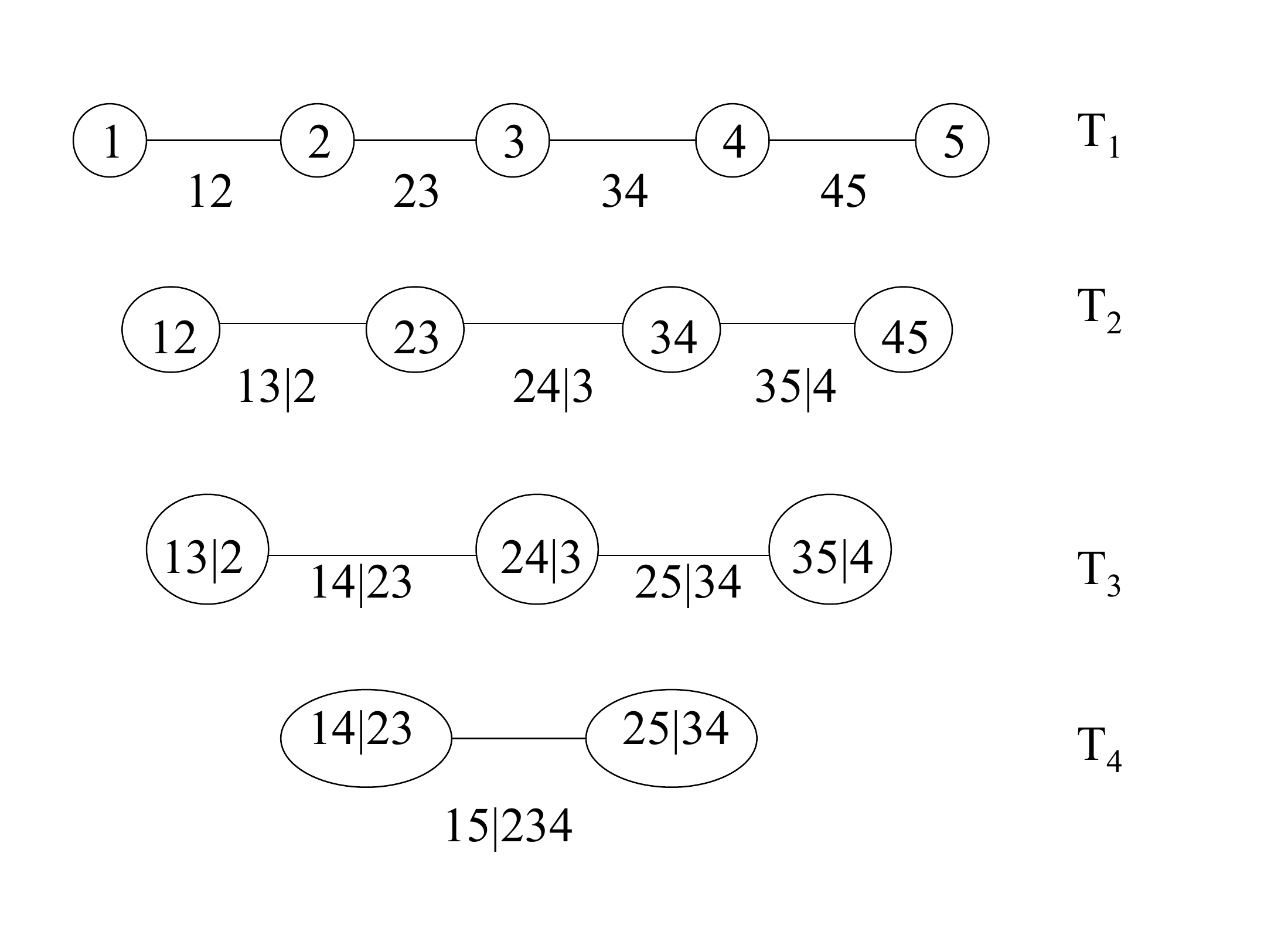}
 \caption{A D-vine with 5 variables, 4 trees and 10 edges. Each edge may be may
 be associated with a pair-copula.}
 \label{fig:DVine}
 \end{figure}

In this paper we use a special case of R-vines called D-vines \cite{Kurowicka04} where each tree is a path. The density $f(x_1,\ldots,x_M)$ corresponding to a D-vine may be written as
\begin{equation}\label{D-Vine}
  f(x_1,\ldots,x_M) = \prod_{j=1}^M f_j(x_j) \times
\end{equation}
\[\prod_{i=1}^{M-1} \prod_{j=1}^{M-i} c_{j,j+i|j+1,\ldots,j+i-1}
\left(F(x_j|x_{j+1},\ldots, x_{j+i-1}), F(x_{j+i}|x_{j+1},\ldots, x_{j+i-1})\right),\]
where index $i$ identifies the trees, and $j$ runs over the edges in each tree. 
The right factor of the righthand side of \eqref{D-Vine} is a product of $M(M-1)/2$ bivariate 
copula densities, and is called an \textit{D-vine copula} density. Note that the arguments of the pair-copulas are 
conditional distributions in all trees except the first, where they are the univariate margins. Figure \ref{fig:DVine} shows a 5-dimensional D-vine with 4 trees and 10 edges.

The density in \eqref{D-Vine} implies a specific order of conditioning. This order can be changed by a simple relabelling of the variables. For example, we can switch the roles of variables $x_1$ and $x_M$. Instead of pair-copulas $c_{1, 2}$ and $c_{M - 1, M}$ we will then get pair-copulas $c_{M, 2}$ and $c_{M - 1, 1}$ in the first tree. Each permutation of $(1, 2, \dots, M)$ therefore gives rise to a different model. These permutations are called \emph{orders} of the D-vine and will play an important role later on.

The key to the construction in \eqref{D-Vine} is that all copulas involved in the decomposition 
are bivariate and can belong to different families. There are no restrictions regarding the copula
types that can be combined; the resulting structure is guaranteed to be valid. A further 
advantage with R-vine copulas is that the conditional distributions $F(x|\bm{v})$ constituting 
the pair-copula arguments can be evaluated using a recursive formula derived in \cite{Joe96}:
\begin{equation}
F(x|\boldsymbol{v})=\frac{\partial C_{xv_j|\boldsymbol{v}_{-j}}(F(x|\boldsymbol{v}_{-j}),F(v_j|\boldsymbol{v}_{-j}))}{\partial F(v_j|\boldsymbol{v}_{-j})}.
\label{eqn:hfunc}
\end{equation}
Here $C_{xv_j|\bm{v}_{-j}}$ is a bivariate copula, $v_j$ is an arbitrary component of $\bm{v}$ and $\bm{v}_{-j}$ 
denotes the vector $\bm{v}$ excluding $v_j$.  By construction, R-vines have the important characteristic that the copulas in question 
are always present in the preceding trees of the structure, so that they are available without extra computations. 

In their general form, vine copulas can represent most continuous multivariate distributions. However, to keep them tractable for inference, the
assumption that the pair copulas 
$$c_{j,j+i|j+1,\ldots,j+i-1}\left(F(x_j|x_{j+1},\ldots, x_{j+i-1}), F(x_{j+i}|x_{j+1},\ldots, x_{j+i-1})\right)$$
are independent of the conditioning variables $x_{j+1},\ldots, x_{j+i-1}$ except through the  conditional marginal distributions, is usually made. This leads
to the so-called \textit{simplified} vine copulas. We will consider both parametric and nonparametric models for the pair-copulas. We use the
\verb|R| package \verb|rvinecopulib| \cite{rvinecopulib} for parameter estimation. More details about parametric and non-parametric estimation
can be found in \cite{Aas07} and \cite{naglerCzado16}, respectively.

\subsection{Shapley contributions: The conditional simulation method} \label{subsec:condsim}

Having determined the multivariate distribution of the explanatory variables, the next step is to compute the contribution function $v(\mathcal{S})$. We propose two different methods for estimating
$v(\mathcal{S})$. In the first, to be described in this section, we generate samples from an estimate of the conditional distribution $f(\bm{x}_{\bar{\mathcal{S}}} | \bm{x}_{\mathcal{S}} = \bm{x}_{\mathcal{S}}^*)$
and use these samples to estimate $v(\mathcal{S})$. In the other, which is treated in Section \ref{subsec:ratio}, $v(\mathcal{S})$ is estimated using ratios of copula densities.

To generate the samples from conditional distributions, we can use the \emph{Rosenblatt transform} \cite{Rosenblatt}
and its inverse. The Rosenblatt transform $\bm{u} = T(\bm{v})$ of a random vector $\bm{v} = (v_1,\ldots,v_M) \sim F$ is defined as
\[u_1= F(v_1),\hspace{0.2cm} v_{2} = F(v_{2}|v_1),\hspace{0.2cm} \ldots, \hspace{0.2cm} u_M =F(v_M|v_1,\ldots,v_{M-1}),\]
where $F(v_m|v_1,\ldots,v_{m-1})$ is the conditional distribution of $v_m$ given $v_1 \ldots, v_{m-1}, \\ m = 2,\ldots,M$. The variables $u_1,\ldots u_M$  are then independent standard uniform variables.
The inverse operation
\[v_1 = F^{-1}(u_1);\hspace{0.2cm} v_{2} = F^{-1}(u_2|u_1);\hspace{0.2cm} \ldots; \hspace{0.2cm} v_M =F^{-1}(u_M|u_1,\ldots,u_{M-1}),\]
can be used to simulate from a distribution. For any joint distribution $F$, if $\bm{u}$ is a vector of independent random variables, $\bm{v} = T^{-1}(\bm{u})$ has distribution $F$. 

In what follows we outline the procedure for generating the $k$th sample from the conditional distribution $F(\bm{x}_{\bar{\mathcal{S}}} | \bm{x}_{\mathcal{S}} = \bm{x}_{\mathcal{S}}^*)$:
\begin{enumerate}
\item For each $j \in \mathcal{S}$, let $u_j^* = \hat{F}_j(x_j^*)$, where $\hat{F}_j$ is the empirical distribution function of $x_j$.
\item Let $\bm{w}_{\bar{\mathcal{S}}}$ be a vector with $|\bar{\mathcal{S}}|$ elements with aritrary values between 0 and 1. Set $\bm{u}=(\bm{w}_{\bar{\mathcal{S}}}, \bm{u}_{\mathcal{S}}^*)$ and 
      let $\bm{v}=T(\bm{u})$, where $T(\cdot)$ is the Roseblatt transform.

\item Generate the vector $\bm{z}_{\bar{\mathcal{S}}}$ by sampling $|\bar{\mathcal{S}}|$ independent uniform U[0,1] distributed variates. 

\item Replace the $|\bar{\mathcal{S}}|$ elements corresponding to the subset $\bar{\mathcal{S}}$ in $\bm{v}$ by $\bm{z}_{\bar{\mathcal{S}}}$.
  
\item Obtain $\bm{u}=T^{-1}(\bm{v})$ using the inverse Rosenblatt transform $T^{-1}(\cdot)$.
  
\item Finally, for each $j \in \bar{\mathcal{S}}$, let $x_j=\hat{F}_j^{-1}(u_j)$, where $\hat{F}_j^{-1}$ is the empirical quantile function of $x_j$.
\end{enumerate}
Step 2 ensures that the values of the conditioning variables are the same in all samples.
Having generated $K$ samples $\bm{x}_{\bar{\mathcal{S}}}^1,\ldots \bm{x}_{\bar{\mathcal{S}}}^K$ from the conditional distribution $f(\bm{x}_{\bar{\mathcal{S}}} | \bm{x}_{\mathcal{S}} = \bm{x}_{\mathcal{S}}^*) $
we use \eqref{eq:montecarl} to compute $v_{\text{KerSHAP}}(\mathcal{S})$.

Whether or not we can compute $T^{-1}(\cdot)$ easily for a given D-vine depends on its \emph{implied sampling orders} \cite{cooke2015}. In particular, the conditioning variables have to appear either first or last in the D-vine structure. For example, a
D-vine with order $1 - 2 - 3 - 4$ allows to easily simulate $\bar{\mathcal S} = \{3, 4\}$ given $\mathcal S = \{1, 2\}$ and $\bar{\mathcal S} = \{1, 2\}$ given $\mathcal S = \{3, 4\}$, but simulating $\bar{\mathcal S} = \{2, 3\}$ given
$\mathcal S = \{1, 4\}$ is only possible through expensive multivariate numerical integration.

More formally, assume that we have a certain permutation $\bm{\pi}=(\pi_1,\ldots \pi_M)$ of $(1, \dots, M)$. The corresponding D-vine may then be used to generate samples from conditional distributions
$f(\bm{x}_{\bar{\mathcal{S}}} | \bm{x}_{\mathcal{S}} = \bm{x}_{\mathcal{S}}^*)$ where $\mathcal S$ either is of the form $\mathcal{S} = \{\pi_1, \dots, \pi_k\}$ or $\mathcal{S} = \{\pi_M, \dots, \pi_{M - k + 1}\}$ for $k = 1, \dots, M$.
In Section \ref{subsec:search}, we use this fact to search for a small set of models that allows for simulation conditionally on any viable coalition $\mathcal S$.

\subsection{Shapley contributions: The ratio method} \label{subsec:ratio}

For vine copula models, conditional simulation often involves numerical integration or inversion, which significantly slows down the algorithms. \cite{nagler2020solving} proposed an alternative way to
approximate conditional expectations based on copulas. The idea is to weight every sample in \eqref{eq:montecarl} in a way that accounts for the dependence. 

It turns out that the appropriate weights are given by a ratio of copula densities. For simplicity denote $u_j = F(x_j)$, $j = 1, \dots, M$. If we have continuous variables, we can compute
\begin{align*}
  v(\mathcal{S}) = \mathbb{E}[g(\bm{x}_{\bar{\mathcal{S}}}, \bm{x}_{\mathcal{S}})| \bm{x}_\mathcal{S} = \bm{x}_\mathcal{S}^*] &= \int g(\bm{x}_{\bar{\mathcal{S}}}, \bm{x}_{\mathcal{S}}^*) f(\bm{x}_{\bar{\mathcal{S}}} | \bm{x}_{\mathcal{S}} = \bm{x}_{\mathcal{S}}^*) \,d\bm{x}_{\bar{\mathcal{S}}}\\
 &= \int g(\bm{x}_{\bar{\mathcal{S}}}, \bm{x}_{\mathcal{S}}^*) f(\bm{x}_{\bar{\mathcal{S}}},\bm{x}_{\mathcal{S}}^*)/f(\bm{x}_{\mathcal{S}}^*) \,d\bm{x}_{\bar{\mathcal{S}}}\\
&= \int g(\bm{x}_{\bar{\mathcal{S}}}, \bm{x}_{\mathcal{S}}^*)\frac{c(\bm{u}_{\bar{\mathcal{S}}},\bm{u}_{\mathcal{S}}^*)\prod_{j=1}^M f(x_j)}{c(\bm{u}_{\mathcal{S}}^*)\prod_{j \in \mathcal{S}}f(x_j)}\,d\bm{x}_{\bar{\mathcal{S}}}\\
&= \int g(\bm{x}_{\bar{\mathcal{S}}}, \bm{x}_{\mathcal{S}}^*)\frac{c(\bm{u}_{\bar{\mathcal{S}}},\bm{u}_{\mathcal{S}}^*)}{c(\bm{u}_{\mathcal{S}}^*)}\frac{f(\bm{x}_{\bar{\mathcal{S}}})}{c(\bm{u}_{\bar{\mathcal{S}}})} \,d\bm{x}_{\bar{\mathcal{S}}}\\
&= E_{\bm{x}_{\bar{\mathcal{S}}}}\left[g(\bm{x}_{\bar{\mathcal{S}}}, \bm{x}_{\mathcal{S}}^*)\frac{c(\bm{u}_{\bar{\mathcal{S}}},\bm{u}_{\mathcal{S}}^*)}{c(\bm{u}_{\bar{\mathcal{S}}})}\right]/c(\bm{u}_{\mathcal{S}}^*)\\
&= E_{\bm{x}_{\bar{\mathcal{S}}}}\left[g(\bm{x}_{\bar{\mathcal{S}}}, \bm{x}_{\mathcal{S}}^*)\frac{c(\bm{u}_{\bar{\mathcal{S}}},\bm{u}_{\mathcal{S}}^*)}{c(\bm{u}_{\bar{\mathcal{S}}})}\right]/E_{\bm{u}_{\bar{\mathcal{S}}}}\left[\frac{c(\bm{u}_{\bar{\mathcal{S}}},\bm{u}_{\mathcal{S}}^*)}{c(\bm{u}_{\bar{\mathcal{S}}})}\right].
\end{align*}
The expression in the third line follows from the definition of a copula given in Section \ref{GaussCop}, while the one in the fourth line is obtained using
\[ f(\bm{x}_{\bar{\mathcal{S}}}) = c(\bm{u}_{\bar{\mathcal{S}}})\prod_{j \in \mathcal{S}}f(x_j).\]
To approximate the last line, we can estimate a vine copula model $\hat c$ and replace the expectations
by a sample average over a (possibly random) subset of the training data:
\begin{equation} \label{eq:ratio}
  v_{\text{KerSHAP}}(\mathcal{S}) = \frac{\sum_{k=1}^K g(\bm{x}_{\bar{\mathcal{S}}}, \bm{x}_{\mathcal{S}}^*)\hat{c}(\bm{u}_{\bar{\mathcal{S}}}^k,\bm{u}_{\mathcal{S}}^*)/\hat{c}(\bm{u}_{\bar{\mathcal{S}}}^k)}
  {\sum_{k=1}^K \hat{c}(\bm{u}_{\bar{\mathcal{S}}}^k,\bm{u}_{\mathcal{S}}^*)/\hat{c}(\bm{u}_{\bar{\mathcal{S}}}^k)}.
\end{equation} 
Similar expressions for discrete or mixed data and theoretical guarantees for \eqref{eq:ratio} can be found in \cite{nagler2020solving}. Note that the formula in \eqref{eq:ratio} is very similar to the one for the empirical method in \cite{aas2019Explaining}.
However, while the weights in that paper were computed using a Gaussian kernel, they are here given as ratios of copula densities. 

The joint vine copula density $\hat{c}(\bm{u}_{\bar{\mathcal{S}}},\bm{u}_{\mathcal{S}}^*)$ is easily computed from \eqref{D-Vine} irrespective of the D-vine order. However, only some of the marginals $\hat{c}(\bm{u}_{\bar{\mathcal{S}}})$ are available in
closed form for a given D-vine. For example, a D-vine with order $1 - 2 - 3 - 4$ allows to easily compute the marginals $c_{1,2}$, $c_{2,3}$, $c_{3,4}$, $c_{1,2,3}$ and $c_{2,3,4}$, but not any other marginals.

To formalize this, we again identify the D-vine structure with a permutation $\bm{\pi}=(\pi_1,\ldots \pi_M)$ of $(1, \dots, M)$. From this permutation we may easily compute all marginals $\hat{c}(\bm{u}_{\bar{\mathcal{S}}})$ where
$\bar{\mathcal S} = \{\pi_k, \pi_{k + 1}, \dots, \pi_\ell\}$ for $1 \le k \le \ell \le M$.


\subsection{Choice of D-vine structures} \label{subsec:search}

We can use D-vine copula models in both the conditional simulation method and the ratio method. Depending on our choice of method, we need to either simulate conditionally from an estimated model or compute a ratio of copula densities. How efficiently we can do this numerically depends on the interplay of the coalition $\mathcal S$ and the order of variables in the D-vine. Generally, there are $M$!/2 distinct D-vines when we have $M$ variables. Usually, when using vines one looks for the D-vine maximising dependence in the first trees. The nature of the problem
treated in this paper is a bit different from the ones previously discussed in the literature.

Let $\mathcal Z$ be the set of all conditional distributions $f(\bm{x}_{\bar{\mathcal{S}}} | \bm{x}_{\mathcal{S}} = \bm{x}_{\mathcal{S}}^*)$ to be used in the conditional simulation method, or all copula marginals $\hat{c}(\bm{u}_{\bar{\mathcal{S}}})$ to be
computed in the ratio method. In the previous two sections, we identified the conditional distributions or copula marginals that may be easily obtained for a given D-vine. In this section we propose a randomized search method that minimizes computational
complexity by finding a small set of D-vine models that covers $\mathcal Z$. The procedure is as follows:
\begin{enumerate}
  \item Generate $B$ random permutations of $(1, \dots, M)$.
  \item For each permutation, find the number of conditional distributions or copula marginals that may be easily obtained (see Sections \ref{subsec:condsim} and \ref{subsec:ratio}).
  \item Pick the permutation that covers most of the remaining sets in $\mathcal{Z}$. Remove the covered sets from $\mathcal Z$.
  \item Go back to step 1 until no subsets are remaining.
\end{enumerate}
The result is a collection D-vine structures such that all conditional distributions/copula marginals may be easily computed. We have used $B=100$ permutations, which gave fairly stable results in our experiments with 10 features. 
Empirically, this approach reduces the number of D-vine models to estimate from $2^M$ to around $2^{M - 2}$ for conditional simulation and to $2^{M - 3}$ for the ratio method. That is, the computational time is
reduced by 75\% -- 87.5\%. 
   
\section{Simulation studies}\label{sec:simulation}

In this section, we discuss a simulation study designed to compare different ways to estimate Shapley values. Specifically, we compare our suggested approaches with \cite{Lundberg}'s independence estimation approach
(below called \textit{independence}) and \cite{aas2019Explaining}'s empirical, Gaussian and Gaussian copula estimation approaches. A short description of each approach is given in Table \ref{tbl:sim_study_outline2}.
For the approaches presented in this paper, we have fitted both a non-parametric and a parametric vine. The independence,
empirical, Gaussian and Gaussian copula approaches are all implemented in the \verb|R| package \verb|shapr| \cite{Sellereite2020}, and the plan is to also include the approaches proposed in this paper.

The simulation model is detailed in Section \ref{burrdist}, the actual design of the experiments is given in Section \ref{subsec:design}. Section \ref{subsec:evaluation} describes the evaluation measure used to quantify the accuracy of the different methods, while
Section \ref{results} gives the results.

\begin{table}[ht]
	\centering
	\begin{tabular}{>{\centering}p{0.2\textwidth} >{\centering}p{0.15\textwidth} >{\arraybackslash}p{0.68\textwidth}}
		\textbf{Method} & \textbf{Citation} & \textbf{Description} \\
		\midrule
          Independence & \cite{Lundberg} & Assume the features are independent. Estimate \eqref{eq:cond_expec} by \eqref{eq:indep}
                                                         where $x_{\bar{\mathcal{S}}}^k$ are sub-samples from the training data set. \\
		\hline
          Empirical & \cite{aas2019Explaining} & Calculate the Mahalanobis distance between the observation being explained and every training instance. Use this distance to calculate a weight for each training instance.
                                                 Approximate \eqref{eq:cond_expec} using a function of these weights. \\
		\hline
          Gaussian  & \cite{aas2019Explaining} & Assume the features are jointly Gaussian.  Sample $N$ times from the corresponding conditional distribution. 
                                                      Estimate \eqref{eq:cond_expec} with \eqref{eq:montecarl} using this sample.\\
          \hline
          Gaussian copula  & \cite{aas2019Explaining} & Assume the dependence structure of the features can be approximated by a Gaussian copula. Sample $N$ times from the corresponding conditional distribution.
                                               Estimate \eqref{eq:cond_expec} with \eqref{eq:montecarl} using this sample.\\
          \hline
	  Parametric cond. sim.       &  & Assume the features are from a vine with all pair-copulas chosen as Clayton Survival copulas. Sample $N$ times from the corresponding conditional distribution.
                                           Estimate \eqref{eq:cond_expec} with \eqref{eq:montecarl} using this sample.\\
          \hline
          Non-parametric cond. sim.   &  & Assume the features are from a non-parametric vine. Sample $N$ times from the corresponding conditional distribution. Estimate \eqref{eq:cond_expec} with \eqref{eq:montecarl} using this sample.\\
          \hline
          Parametric ratio    &  & Assume the features are from a vine with all pair-copulas chosen as Clayton Survival copulas. Estimate \eqref{eq:cond_expec} with \eqref{eq:ratio}.\\
          \hline
          Non-parametric ratio&  & Assume the features are from a non-parametric vine. Estimate \eqref{eq:cond_expec} with \eqref{eq:ratio}.\\
		\midrule
	\end{tabular}
	\vspace{0.2cm}
	\caption{A short description of the approaches used to estimate \eqref{eq:cond_expec} in the simulation studies.}
	\label{tbl:sim_study_outline2}
\end{table} 

\subsection{Simulation model}\label{burrdist}

To evaluate the different approaches, we need cases for which we know the true feature distribution. Moreover, we have to use multivariate distributions that have known conditional distributions. There are not many such
distributions, but one example, which allows for heavy-tailed and skewed marginals and non-linear dependence, is the multivariate Burr distribution. 

The $M$-dimensional Burr distribution has the density \cite{Takahasi1965}
\[ f_M(\bm{x}) = \frac{\Gamma(p+M)}{\Gamma(p)}\left(\prod_{m=1}b_m\,r_m\right)\frac{\prod_{m=1}^M x_m^{b_m-1}}{\left(1+\sum_{m=1}^M r_m\,x_m^{b_m}\right)^{p+M}},\]
for $x_m > 0$. Here, $p$, $b_1,\ldots,b_M$ and $r_1,\ldots r_M$ are the parameters of the distribution. The Burr distribution is a compound Weibull distribution with the gamma distribution as compounder \cite{Takahasi1965}.
It can be regarded as a special case of the Pareto IV distribution \cite{YARI2006}.

Any conditional distribution of the multivariate Burr distribution is also a multivariate Burr distribution \cite{Takahasi1965}. The conditional density $f(x_1,\ldots,x_S|x_{S+1}=x_{S+1}^*,\ldots,x_M=x_{M}^*)$ is an
$S$-dimensional Burr density with parameters $\tilde p$, $\tilde b_1, \dots, \tilde b_S$, $\tilde r_1, \dots, \tilde r_S$, where $\tilde p = p + M - S$ and for all $j = 1, \dots, S$, 
\begin{align*}
  \tilde b_j = b_j, \qquad \tilde r_j = \frac{r_j}{1+\sum_{m=S+1}^M r_m\,\left(x_m^*\right)^{b_m}}.
\end{align*}

According to \cite{Cook81}, the copula corresponding to the multivariate Burr distribution is a Clayton survival copula. Thus, the multivariate Burr distribution may be represented by a vine copula where
the pair-copulas are bivariate Clayton survival copulas. For this reason we have fitted both a non-parametric and a parametric vine for the two approaches presented in this paper. For the parametric vines
the models are correctly specified in this simulation example. Hence, if the non-parametric vines have similar performance to the corresponding parametric vines, it indicates that the non-parametric vines provide a
satisfactory fit to the multivariate Burr distribution.

In our experiments, we simulate data from 3 different 10 dimensional Burr distributions.  All three distributions have 
\begin{align}
\boldsymbol{b} &= (2,4,6,2,4,6,2,4,6,6)\notag \\
\boldsymbol{r} &= (1,3,5,1,3,5,1,3,5,5), \notag 
\end{align}
while they have $p$ equal to 0.5, 1, and 1.5, respectively. The three values of $p$ correspond to a pair-wise Kendall's $\tau$ of 0.5, 0.33, and 0.25, respectively.

In addition to the feature distribution, we need to specify the sampling model for the response $y$ and the machine learning approach used to fit the predictive model $g(\bm{x})$.
Inspired by \cite{Chang2019} we chose the following non-linear and heteroscedastic function for $y$:
\begin{equation}\label{yeq}
  y = u_1\,u_2\exp(1.8\,u_3\,u_4) + u_5\,u_6\exp(1.8\,u_7\,u_8) + u_9\exp(1.8\,u_{10})+0.5(u_1+u_5+u_9)\epsilon.
\end{equation}
Here $\bm{x}$ is multivariate Burr distributed and $u_m=F_m(x_m)$ where $F_m(\cdot)$ is the true parametric distribution function. Finally, $\epsilon$ is standard normal distributed
and independent of all the $x_m$s. 

\subsection{Experimental design} \label{subsec:design}

We perform 3 different experiments with training sample sizes $N_{\mbox{train}}$ equal to 100, 1,000 and 10,000, respectively. In each experiment we repeat the following steps 50 times for 
each of the 3 different Burr distributions described in Section \ref{burrdist}:
\begin{enumerate}
\item Sample $N_{\mbox{train}}$ training observations from the chosen Burr distribution
\item Compute the corresponding $y$ values using \eqref{yeq}.
\item Fit a Random forest with 500 trees using the R package {\tt ranger} \cite{ranger} with default parameter settings to the training data. 
\item Sample $N_{\mbox{test}}=100$ test observations from the chosen distribution.
\item For all possible subsets $\mathcal{S}$ and all test observations $\bm{x}^*$:
    \begin{itemize}
      \item If one of the ratio methods: Compute $v_{\text{KerSHAP}}(\mathcal{S})$ using \eqref{eq:ratio} with $K=1,000$.
      \item If the empirical method: Compute $v_{\text{KerSHAP}}(\mathcal{S})$ using the formula given in Section 3.3 in \cite{aas2019Explaining} with $\eta=0.95$.
      \item If one of the remaining methods in Table \ref{tbl:sim_study_outline2}: Generate $K=1,000$ samples from the conditional distribution $p(\bm{x}_{\bar{\mathcal{S}}} | \bm{x}_{\mathcal{S}} = \bm{x}_{\mathcal{S}}^*)$ and
        compute $v_{\text{KerSHAP}}(\mathcal{S})$ using \eqref{eq:montecarl}. 
      
      \end{itemize}
  \item For all test observations $\bm{x}^*$, compute the Shapley value using \eqref{eq:shapley} with the $v_{\text{KerSHAP}}(\mathcal{S})$ values for all subsets $\mathcal{S}$ for the current test observation.
\end{enumerate}
For all approaches, the multivariate model for the features is fitted using the training data.

\subsection{Evaluation method}\label{subsec:evaluation}

We measure the performance of each method based on the mean absolute error (MAE), across both the features and the sample space. MAE is defined as 
\begin{equation}\label{eq:MAE_cat}
\text{MAE}(\text{method } q) = \frac{1}{T} \sum_{i=1}^T \frac{1}{M} \sum_{j=1}^{M}|\phi_{j, \text{true}}(\bm{x}_i) - \phi_{j, q}(\bm{x}_i) |,
\end{equation}
where $\phi_{j, q}(\bm{x})$ and $\phi_{j, \text{true}}(\bm{x})$ denote, respectively, the Shapley value estimated with method $q$ and the corresponding true Shapley value for the prediction $g(\bm{x})$.  Further,
$M$ is the number of features and $T$ is the number of test observations.  We use Monto Carlo integration with 10,000 samples to compute the exact Shapley values.

As stated in Section \ref{subsec:design},  for each feature distribution and choice of $N_{train}$ we repeat the test procedure 50 times and report the average MAE over those 50 repetitions. Hence, 
the quality of the Shapley values is evaluated based on a total of $T=5,000$ test observations. Sampling new data for each batch reduces the influence of the exact shape of the fitted predictive model.

\subsection{Results} \label{results}

\begin{figure}[t!]
  \centering 
  \includegraphics[width = \textwidth]{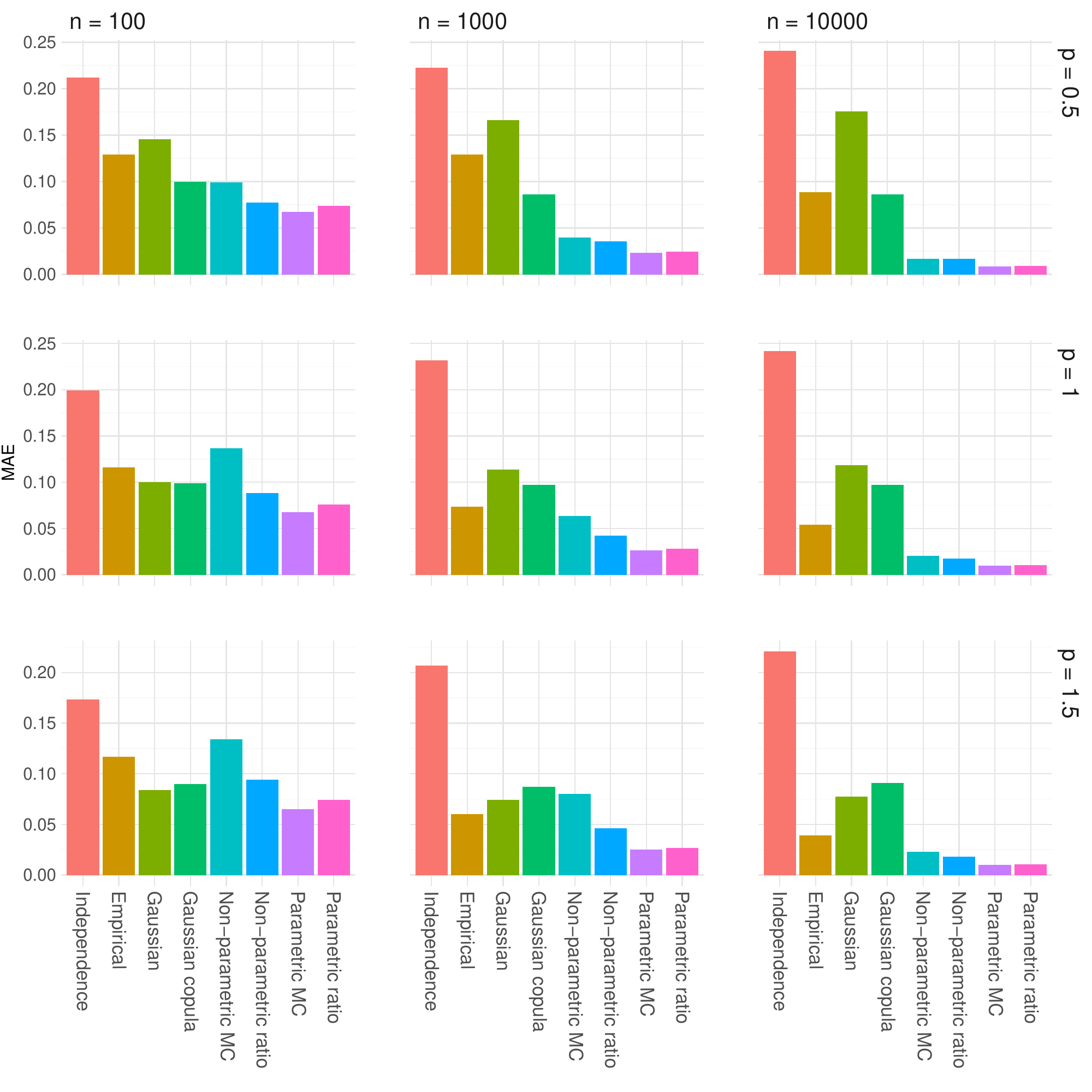}
  \vspace{0.2cm}
  \caption{MAE for each combination of sample size, Burr distribution parameters and method. Each MAE-value is computed from 5,000 test observations. The values  0.5, 1, and 1.5 of $p$ correspond to pair-wise Kendall's $\tau$s of 0.5, 0.33, and 0.25,
    respectively.}
  \label{sim_results}
\end{figure}

The results of the simulation study are shown in Figure~\ref{sim_results}. The nine panels correspond to different combinations of sample size $N_{\mbox{train}}$ (columns) and dependence parameter $p$ (rows). Each bar represents the
MAE achieved by a particular method, where smaller values indicate higher accuracy.

Analogous to \cite{aas2019Explaining}, we clearly see that the independence method is not suitable for estimating Shapley values when covariates are dependent. The other methods have more similar performance, but with the vine-based methods favored overall.
The parametric vine methods perform slightly better than the non-parametric ones in all scenarios. This is to be expected, because, as previously stated, the true simulation model can be represented as a vine copula with
survival Clayton pair-copulas. Hence, the parametric models are correctly specified. On real data sets, the parametric assumption will rarely hold and can be severly violated, however.

For very small sample sizes ($N_{\mbox{train}}= 100$), even the (correctly specified) parametric vine methods have only a small advantage and the non-parametric methods are outperformed by some of their competitors when $p=1.5$
(weaker dependence). This is not surprising, since we are estimating very complex models  --- up to 90 parameters for parametric vines --- from very limited information. For medium to large samples ($N_{\mbox{train}} \ge 1\,000$),
the vine-based methods outperform their competitors by a decent margin. When the dependence is strong ($p=0.5$), the MAE of the non-parametric ratio method is only approximately 20\% of the MAE of the best of the previously proposed methods
for $N_{\mbox{train}}= 10,000$ and the corresponding ratio for $p=1.5$ is 50\%.

It also becomes apparent that the non-parametric ratio method performs slightly better than its Monte Carlo analogue. The likely reason is that conditional simulation involves many numerical approximations for integration and inversion
that accumulate. The two parametric vine methods on the other hand have virtually the same accuracy.


\begin{figure}[t!]
  \centering 
  \includegraphics[width = \textwidth]{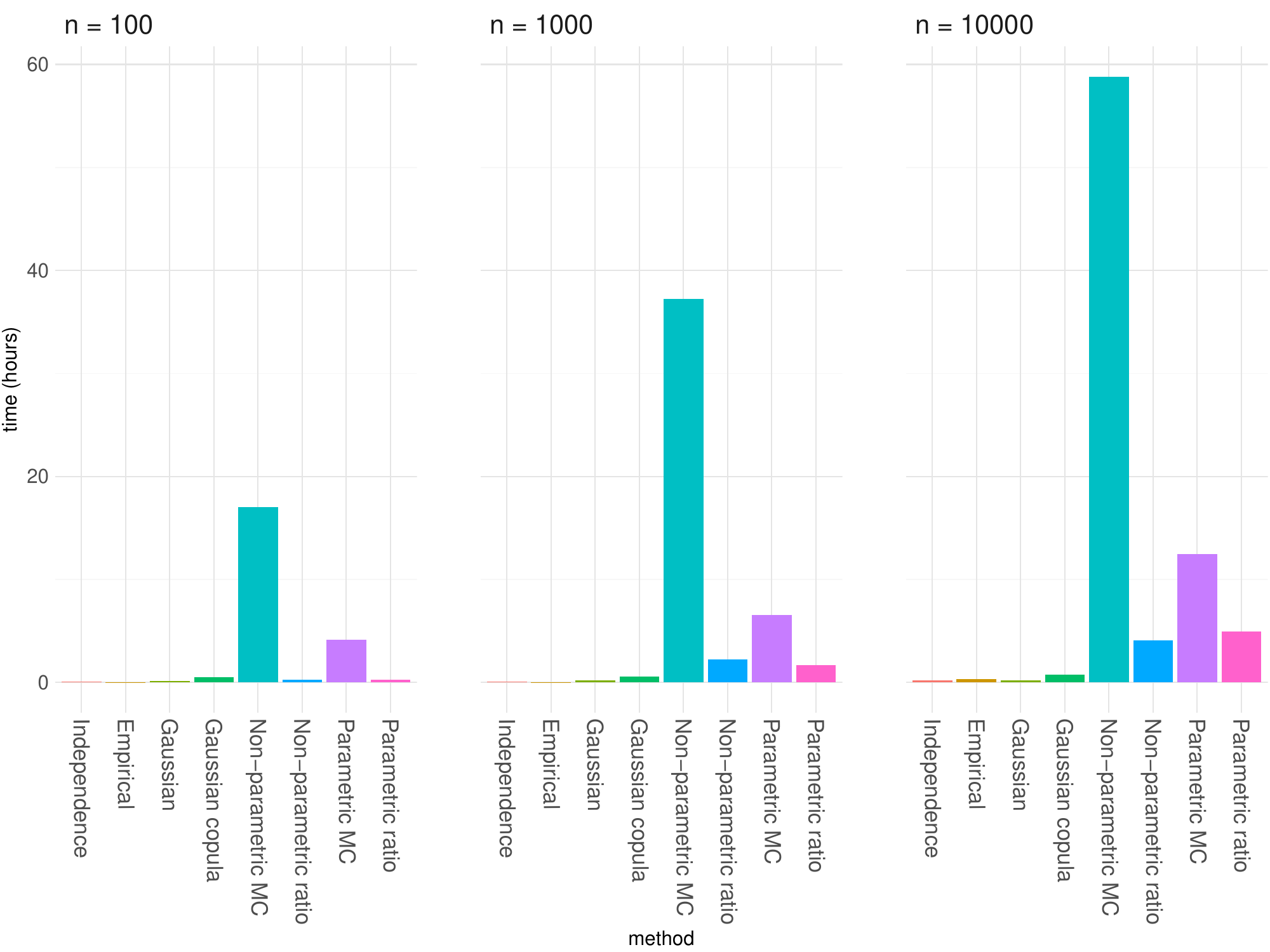}
  \vspace{0.2cm}
  \caption{Average computation time (CPU hours) required for each sample size, method, on  100 test observations.}
  \label{sim_time}
\end{figure}

We can conclude that our vine-based methods improve over previous methods for estimating Shapley values. However, that comes at a computational cost. Figure~\ref{sim_time} shows the average computation time required to estimate Shapley values
for 100 test observations (including fitting copula models). We can confirm that the vine based methods are slower than its competitors. The ratio methods are around 10x slower than the Gaussian copula method, and the Monte Carlo methods even up to 100x.
For practical purposes, the ratio method is therefore preferred. We also note that computation times are generally large. This is mainly due to the fact that we have to compute a large number of different Shapley contributions
$v(\mathcal S)$ ($2^{10} = 1,024$ for 10 covariates). This can be mitigated substantially by parallelizing computations and/or using the approximated weighted least squares method proposed by \cite{Lundberg}. The latter approach, which is
thoroughly described in Section 2.3.1 in \cite{aas2019Explaining} requires only a subset of Shapley contributions to be computed.

\section{Real data example}\label{sec:data}
In this section, we apply the methods discussed in this paper on the Abalone data set (available at http://archive.ics.uci.edu/ml/datasets/Abalone). It has previously been used in several
machine learning studies, see e.g. \cite{Sahin2018, Smith2019}. Moreover, it has been used in the related vine copula studies \cite{Haff2016, Chang2019, CzadoBook19}.  The data
originate from a study by the Tasmanian Aquaculture and Fisheries Institute. An abalone is a kind of edible sea snail,
the harvest of which is subject to quotas. These quotas are based partly on the age distribution of the abalones. To determine an abalone´s age, one cuts the shell through
the cone, stains it, and counts the number of rings through a microscope. This is
a highly time-consuming task. Hence, one would like to predict the age based on physical measurements that are easier to obtain. The Abalone data set was originally used for this purpose.
It consists of 4,177 samples on the following 9 variables: {\tt Sex}, {\tt Length}, {\tt Diameter}, {\tt Height}, {\tt Whole weight}, {\tt Shucked weight}, {\tt Viscera weight},
{\tt Shell weight} and {\tt Age} measured by number of rings.

We do not include the variable {\tt Sex} in our study since it is a discrete variable. Note that the use of regular vines does not exclude discrete data; examples of discrete and mixed discrete vines
may be found for instance in \cite{panagiotelis2012} and \cite{stober2015}. However, many of the methods become more complicated when discrete data are involved.

\begin{figure}[t!]
	\begin{center}
          \includegraphics[width=1.0\linewidth]{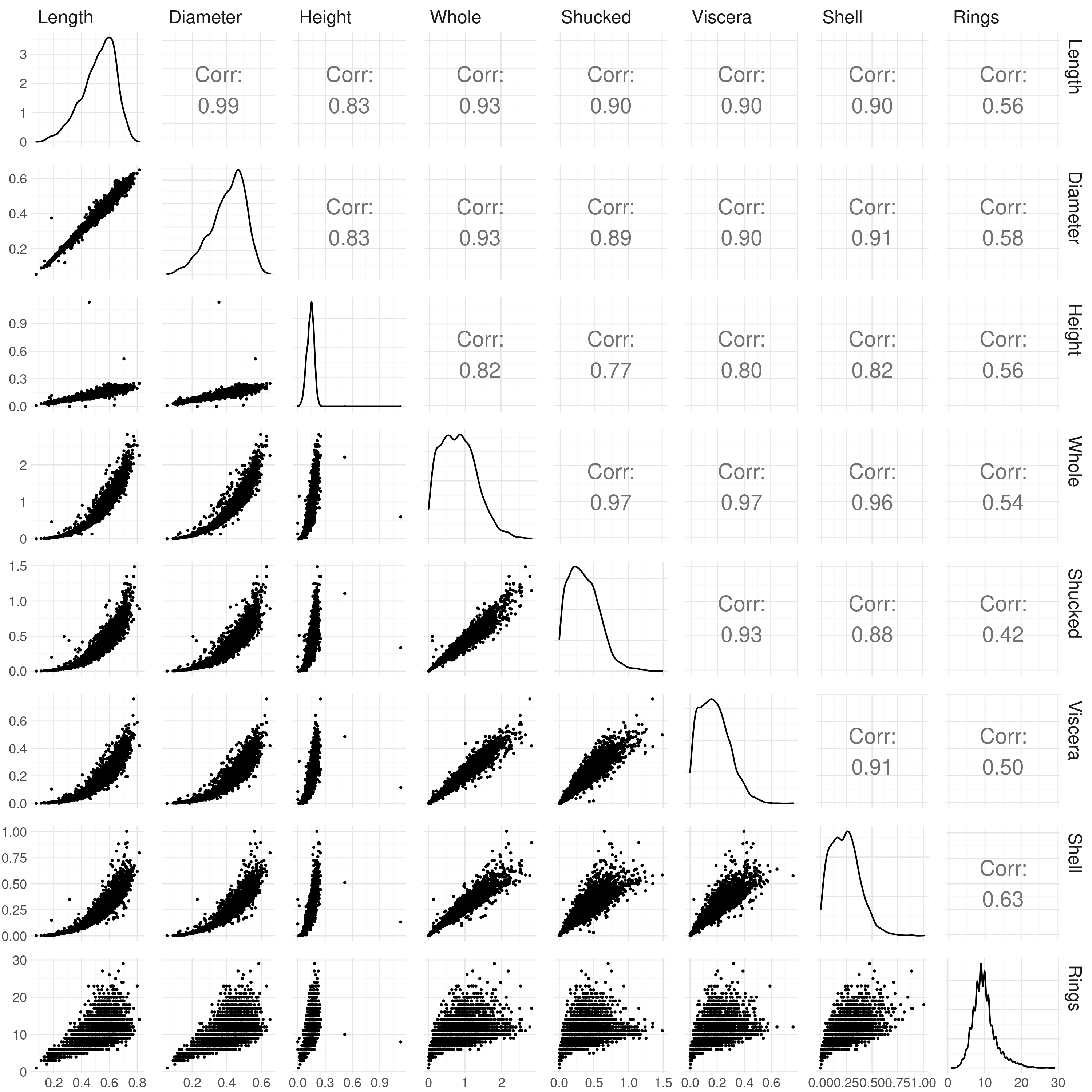}
	\end{center}
	\caption{Pairwise scatter plots, marginal density functions and pairwise correlation coefficients for the explanatory variables and the response variable.  
		\label{fig:hist}}
            \end{figure}

Figure \ref{fig:hist} shows the pairwise scatter plots, marginal density functions and pairwise Pearson correlation coefficients. There is clear non-linearity and heteroscedaticity among the pairs
of variables. Moreover, it can be noted that all pairwise correlations between the explanatory variables are higher than 0.775.

We treat the age prediction as a regression problem.
The Abalone data set was divided into a training set and a test set, containing
4,077 and 100 observations, respectively. We fitted a Random forest model with 500 trees to the training data, using the R package {\tt ranger} with default parameter
settings. Then, this model was used to predict the age (number of rings) for the observations in the test data set.

\begin{figure}[t!]
\begin{center}
  \includegraphics[width=1.0\linewidth]{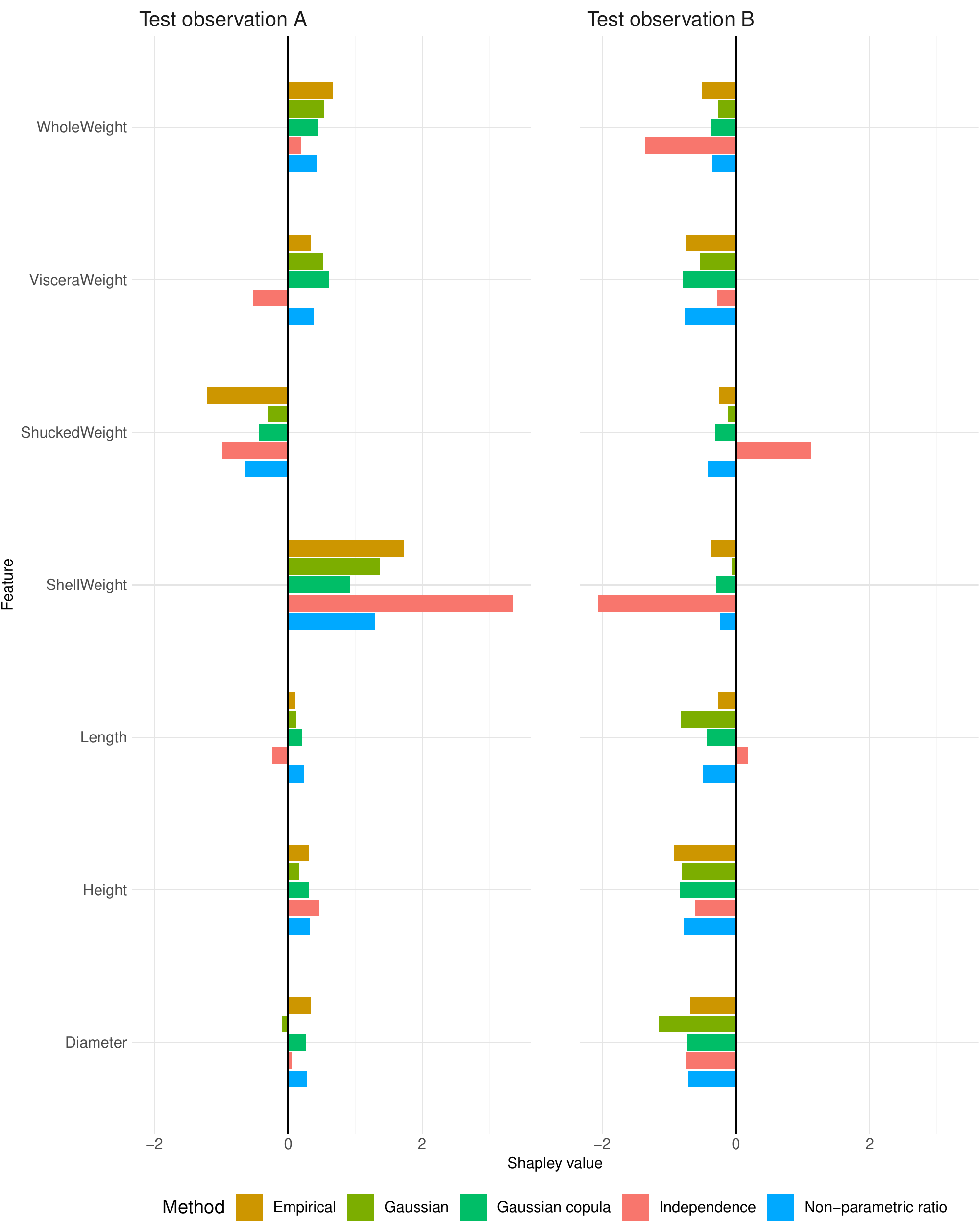}
	\caption{Shapley values for two of the test observations in the real data set computed using the different methods.
		\label{fig:shapleyValues}}
	\end{center}
\end{figure}

Since the non-parametric ratio method was the fastest, most stable, and best performing of the new vine copula methods in the simulation study, we compare the performance of
this method with the independence, empirical, Gaussian and Gaussian copula approaches. Figure \ref{fig:shapleyValues} shows the Shapley values for two of the test observations. The Shapley values computed by the different methods are quite different.
As expected, the independence method is the one that differs most from the other ones. Take e.g. test observation B.  Here, the Shapley values for all variables computed using the non-parametric ratio method have
the same sign and they are of quite similar magnitude.
For the independence approach, however, the Shapley values show much larger variation, both when it comes to sign and magnitude. The Shapley values produced by the Gaussian copula method are the ones that are most similar to those
produced by the non-parametric ratio method. This is as expected, since this method is the one that bears most resemblance with the vines methods. However, we observe some differences, e.g. for the variables
{\tt ShuckedWeight} and {\tt Shell Weight} for test observation A.

A problem with evaluating Shapley values for real data is that there is no ground truth. Hence, we have to justify the results in other ways. In what follows, we use mainly the same framework as that proposed in \cite{aas2019Explaining}.
For all approches treated in this paper, the Shapley value is a weighted sum of differences $v(\mathcal{S}\cup \{j\})-v(\mathcal{S})$ for several subsets $\mathcal{S}$. However, the approaches differ in how $v(\mathcal{S})$, or more specifically,
the conditional distribution  $p(\boldsymbol{x}_{\bar{\mathcal{S}}}|\boldsymbol{x}_{\mathcal{S}}=\boldsymbol{x}_{\mathcal{S}}^*)$, is estimated. Hence, if we are able to show that the samples from the conditional distributions
generated using the non-parametric method are more representative than the samples generated using the previously proposed methods, it is likely that the Shapley values obtained using the non-parametric method are the
most accurate.

\begin{figure}[t!]
	\begin{center}
          \includegraphics[width=1.1\linewidth]{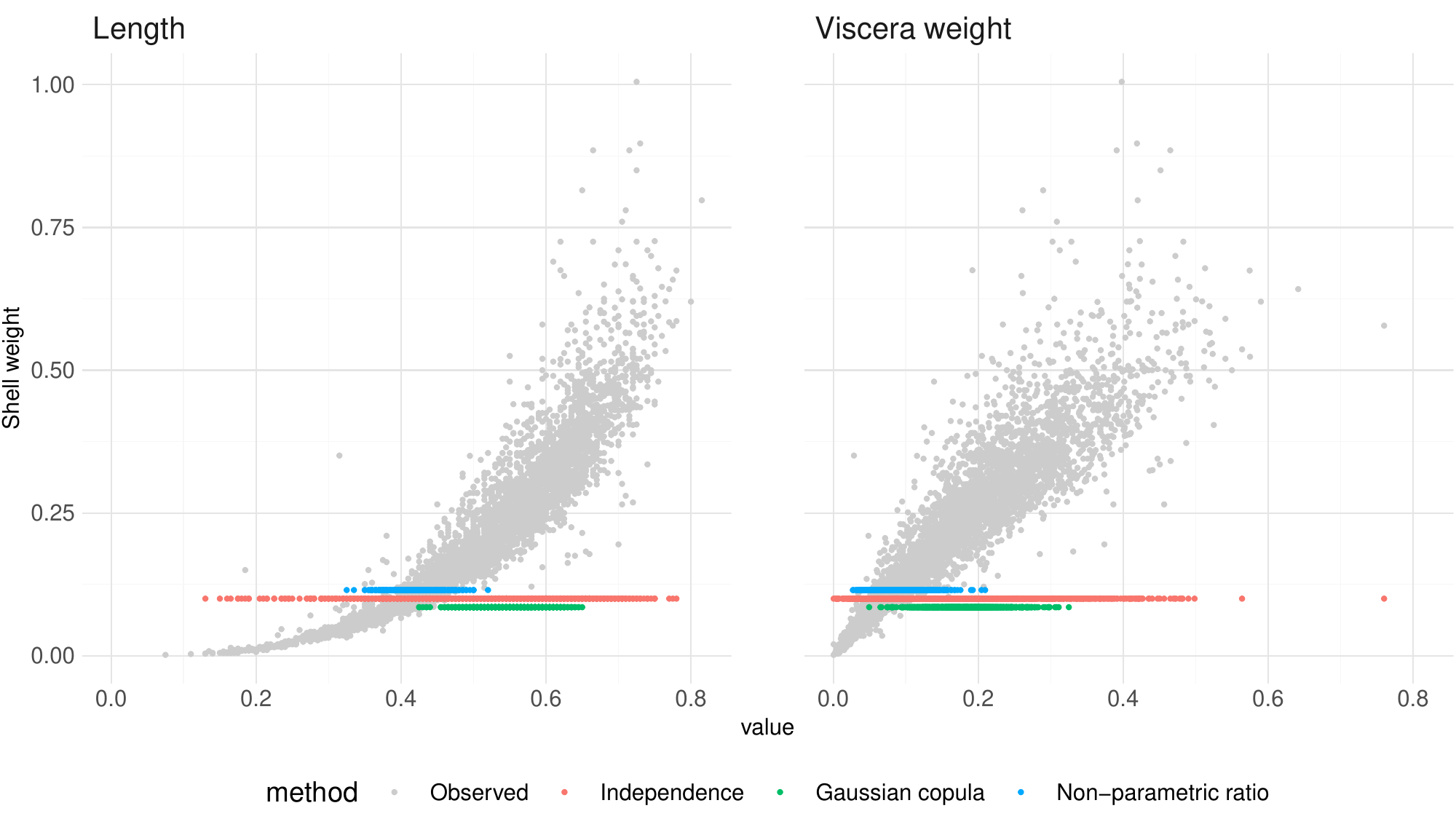}
	\end{center}
	\caption{{\tt Length} against {\tt Shell weight} (left) and {\tt Viscera weight} against {\tt Shell weight} (right). The grey dots are the training data. The blue dots are the samples from the
          conditional distribution of the variable at the x-axis given that {\tt Shell weight} is equal to 0.1 generated using the non-parametric ratio method, the green dots are the corresponding samples
          generated using the Gaussian copula method, and the red dots are the samples generated using the independence method. Note that the red and green dots have been slightly displaced vertically to
          improve visibility of the figure.
         \label{fig:V4}}
         \end{figure}

Since there are many conditional distributions involved in the Shapley formula, we will not show all here. However, we have included some examples that illustrate that the non-parametric method gives more
correct approximations to the true conditional distributions than the other approaches. First, Figure \ref{fig:V4} shows plots of {\tt Length} against {\tt Shell weight} and {\tt Viscera weight} against
{\tt Shell weight}.
The grey dots are the training data. The blue dots are the samples from the conditional distribution of the variable at the x-axis given that {\tt Shell weight} is equal to 0.1, generated using our method. 
The green and red dots are the corresponding samples generated using the Gaussian copula and independence approaches, respectively. It should be noted that the non-parametric ratio method does not involve any simulation.
However, the method has an implicit statistical model that we can sample from for illustrative purposes. \eqref{eq:ratio} can be seen as an expectation $E_P[g(\bm x_{\bar{\mathcal{S}}}, \bm x_{\mathcal{S}^*})]$ with
respect to a model $P$ that assigns probability
\begin{align*}
  \pi_\ell =  \frac{\hat c(\bm u_{\bar{\mathcal{S}}}^\ell, \bm u_{\mathcal{S}}^*) / \hat c(\bm u_{\bar{\mathcal{S}}}^k)}{\sum_{k = 1}^K \hat c(\bm u_{\bar{\mathcal{S}}}^k, \bm u_{\mathcal{S}}^*) / \hat c(\bm u_{\bar{\mathcal{S}}}^k)}
\end{align*}
to the $\ell$th observation $\bm x^\ell$. Hence, we can simulate from this implicit model by drawing with replacement from the original observations $\bm x^1, \dots, \bm x^K$ using $(\pi_1, \dots, \pi_K)$ as sampling probabilities.
Both the Gaussian copula approach and the independence approach generate samples that are unrealistic, in the sense that they are far outside the range of what is observed in the training data.
It might be confusing that this is the case for the Gaussian copula. This is due to the fact that we are sampling in the lower tail, where there is very strong tail-dependence that the Gaussian copula is missing out on.
It is well known that evaluation of predictive machine learning models far from the domain at which
they have been trained, can lead to spurious predictions. Thus, it is important that the explanation methods are evaluating the predictive model at appropriate feature combinations. 
The samples generated by the ratio method are inside the range of what is observed in the training data. 

\begin{figure}[t!]
	\begin{center}
		\includegraphics[width=0.9\linewidth]{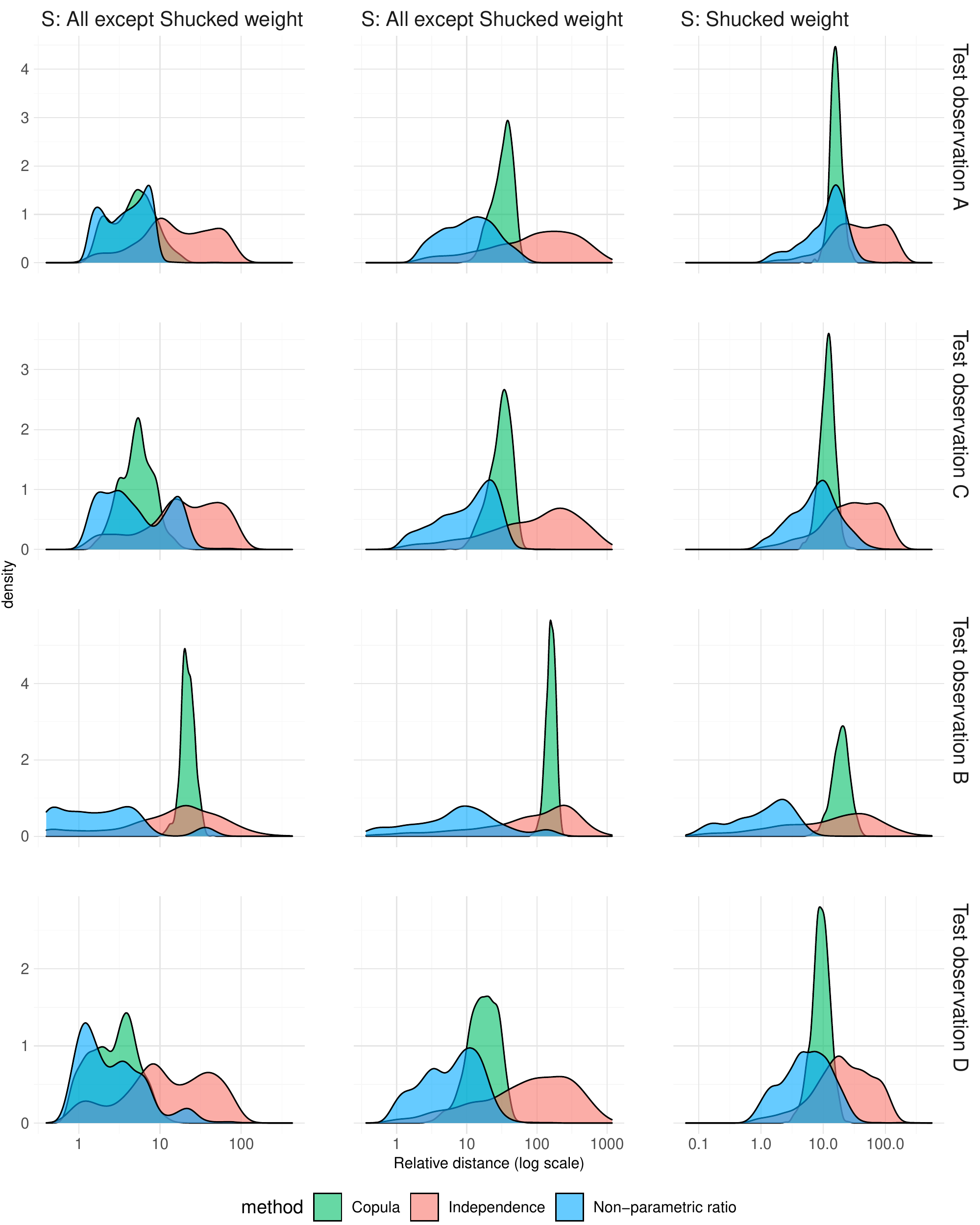}
	\end{center}
	\caption{Probability densities of mean Mahalanobis distances for three different conditional distributions and three different individuals. See the text for a further description.
		\label{threeCondDist}}
            \end{figure}

In Figure \ref{threeCondDist} we study three different conditional distributions involved in the Shapley formula:
\begin{itemize}
\item The conditional distribution of {\tt Shell weight} given all the other variables.
\item The conditional distribution of {\tt Length} and {\tt Shucked weight} given {\tt Viscera weight} and {\tt Shell weight}.
\item The conditional distribution of all variables except {\tt Shucked weight} given {\tt Shucked weight}.
\end{itemize}  
For all the three distributions, we generate 1,000 samples for three of the test observations using the non-parametric ratio, Gaussian copula and independence approaches. That is, we condition on four different sets of values.
For each combination of test observation, conditional distribution and method, we compute the mean Mahalanobis distance between each sample and its ten nearest training samples, resulting in 1,000 different mean
distances. Each panel of Figure \ref{threeCondDist} shows the probability densities of such mean distances for a specific test observation and a specific conditional distribution (test observations A and B
are the same as those in Figure \ref{fig:shapleyValues}).
If the generated samples are realistic, we would expect the majority of the mean distances to be small.

For all conditional distributions and all test observations, the mode of the density corresponding to the independence approach is larger than those of the two other densities,
indicating that the samples generated by the Gaussian copula and non-parametric ratio approaches are more realistic than those generated by the independence approach. Further, for the majority of the test observations/conditional
distributions, the Mahalanobis distances corresponding to the non-parametric ratio approach are smaller than those corresponding to the independence and Gaussian copula approaches. 

To summarize, we have illustrated that the Shapley values computed using the non-parametric ratio method and the previously proposed methods are different. We have tried to justify that this is because
the non-parametric ratio method gives more correct approximations to the true conditional distributions for this data set.

\section{Summary and discussion}\label{sec:conclusion}
Shapley values are a model-agnostic method for explaining individual predictions with a solid theoretical foundation. The original development of Shapley values for prediction explanation relied on the assumption that the features
being described were independent. If the features in reality are dependent this may lead to incorrect explanations. Hence, there have recently been attempts of appropriately modelling/estimating the dependence between
the features. Although the proposed methods clearly outperform the traditional approach assuming independence, they have their weaknesses. In this paper we have proposed two new approaches for modelling the dependence between
the features. Both approaches are based on vine copulas, which are flexible tools for multivariate non-Gaussian distributions able to characterise a wide range of complex dependencies.

We have performed a comprehensive simulation study, showing that our approaches outperform the previously proposed methods. We have also applied the different approaches
to a real data set, where the predictions to be explained were produced by a Random forest classifier designed to predict the age of an abalone (sea snail). In this case the true Shapley values are not known, but we provide results which indicate
that the vine based approaches provide more sensible approximations than the previously proposed methods.

The main part of the methodology proposed in this paper may be used for many other applications than computing Shapley values. The need for expressing statistical inference in terms of conditional quantities is ubiquitous
in most natural and social sciences \cite{Otneim2018}. An obvious example is the estimation of the mean of some set of response variables conditioned on sets of explanatory variables taking specified values \cite{Chang2019}.
Other common tasks are the forecasting of volatilities or quantiles of financial time series conditioned on past history \cite{Schittenkopf2000}. Problems of this kind often call for some sort of regression analysis, like the one presented in this paper.

The challenging issue in conditional density estimation is to circumvent the curse of dimensionality. Several methods have been proposed to estimate conditional densities;
the classical kernel estimator \cite{rosenblatt1956},  which has been refined and developed in many directions, see for example \cite{Holmes2010, bertin2016, izbicki2017,Nguyen2018}; local polynomial estimators \cite{Hyndman1996,Fan1996},
and a local Gaussian correlation estimator \cite{Otneim2018}. However, most of these methods, if not all, are computationally intractable when either $\boldsymbol{x}_{\mathcal{S}}$ or $\boldsymbol{x}_{\bar{\mathcal{S}}}$ are not
univariate, or both have dimension above 3-4. The vine-based approaches proposed in this paper work well when both $\boldsymbol{x}_{\mathcal{S}}$ or $\boldsymbol{x}_{\bar{\mathcal{S}}}$ are high dimensional. Hence, the methodology
proposed in this paper may be regarded as a contribution to the field of non-parametric conditional density estimation.

\section*{Acknowledgements}
This work is supported by the Norwegian Research Council grant 237718.



\bibliographystyle{splncs04}

\bibliography{references}

\end{document}